\documentclass[pre,twocolumn,preprintnumbers,amsmath,amssymb,superscriptaddress,reprint]{revtex4} % reprint option for length guidance

\usepackage{graphicx}% Include figure files
\usepackage{dcolumn}% Align table columns on decimal point
\usepackage{bm,amsmath}% bold math
\usepackage{here}% bold math

\renewcommand{\vec}[1]{{\bm #1}}

\begin{document}
\title{Response function of turbulence computed via fluctuation-response relation of a Langevin system with vanishing noise}
\author{Takeshi \surname{Matsumoto}}
\email{takeshi@kyoryu.scphys.kyoto-u.ac.jp}
\affiliation{%
Division of Physics and Astronomy,
Graduate School of Science,
Kyoto University,
Kyoto, 606-8502, Japan}
\author{Michio \surname{Otsuki}}
\affiliation{% 
Department of Materials Science,
Shimane  University, 
Matsue, 690-8504, Japan}
\author{\surname{Ooshida} Takeshi}
\affiliation{%
Department of Mechanical and Aerospace Engineering,
Tottori University, Tottori, 680-8552, Japan}
\author{Susumu \surname{Goto}}
\affiliation{%
Graduate School of Engineering Science,  
Osaka University, Toyonaka, 560-8531, Japan}
\author{Akio \surname{Nakahara}}
\affiliation{%
Laboratory of Physics, College of Science and Technology,
Nihon University, Funabashi, 274-8501, Japan} 
\date{\today}
\begin{abstract}
For a shell model of the fully developed turbulence
and the incompressible Navier-Stokes equations in the Fourier space,
when a Gaussian white noise is artificially added to the equation of each mode,
an expression of the mean linear response function in terms of the velocity correlation functions 
is derived by applying the method developed for nonequilibrium Langevin systems
[Harada and Sasa, Phys.~Rev.~Lett. \textbf{95}, 130602 (2005)]. 
We verify numerically for the shell model case that the derived expression of the response function,
as the noise tends to zero, converges to the response function of the noiseless shell model. 
\end{abstract}

\maketitle
\textit{Introduction}~
Tools of statistical mechanics are indispensable for research  
of fluid turbulence. We here focus on relation between
the linear response function and the correlation function,
known as fluctuation-response relation (FRR).
The simplest form of the FRR is realized in thermally equilibrium systems:
the linear response function is proportional to the autocorrelation function
of a dynamical variable, with the proportional constant being the inverse 
temperature.
This classical FRR does not hold in general 
for a driven dissipative system with a non-Gaussian distribution function
in a nonequilibrium steady state  as reviewed in \cite{mprv08}.
For statistically steady-state, homogeneous and isotropic turbulence,
researchers have asked the questions: (1) what kind of FRR holds?
(2) can its linear response function be expressed in terms of
velocity correlation functions?

The FRR of turbulence was studied first by Kraichnan
in his influential closure approximation, known as the direct-interaction
approximation (DIA) \cite{k59} (see also \cite{vt}). 
A major goal of statistical theories
is to derive the Kolmogorov energy spectrum $E(k) \propto k^{-5/3}$ from 
the incompressible Navier-Stokes equations \cite{f, ga}. 
The energy spectrum $E(k)$ is the  average of the equal-time autocorrelation 
function of the velocity Fourier modes on the spherical surface with radius 
$k$ in the wavenumber space. To obtain a closure for $E(k)$,
Kraichnan considered the mean linear response function in DIA.
In the latest versions of DIA in the Lagrangian frame of
reference \cite{lra,kg}, successfully reproducing the $k^{-5/3}$ spectrum,
the autocorrelation function and the response function 
are proportional as a result of the closure approximation.
A direct numerical simulation result, though available for the Eulerian frame only,
indicates that
the correlation function and the response function are not proportional 
at the Kolmogorov dissipation scale with moderate Reynolds numbers \cite{cq}.
For a dynamical system model of turbulence in the Lagrangian frame, known as the Gledzer-Ohkitani-Yamada (GOY)
shell model \cite{g,oy,br}, the FRR is numerically studied in \cite{bdlv}, 
demonstrating that the proportionality does not hold for the shell variables
in the inertial range. This is consistent with the strong non-Gaussianity 
of the shell variables.
We here consider expressions of the response functions of the last two cases
in a unified manner using a recent result of nonequilibrium statistical mechanics.
The result we rely on is established by Harada and Sasa ~\cite{hs05, hs06}
to derive a general FRR for a class of nonlinear Langevin systems,
which has been verified experimentally in a thermally activated system since \cite{tjnms}
(for FRR in a deterministic setting, see, e.g., \cite{ts, coh, ho}).

Obviously the macroscopic fluid dynamical system describing turbulence, 
where the thermal driving is unnecessary,
is different from the nonlinear Langevin system.
Nevertheless there appears to be a simple way to bridge the two systems: 
we formally add the Gaussian white noise
to the fluid dynamical equations without worrying 
about its physical origin; next we derive various relations
with the powerful weaponry of the stochastic systems \cite{se};
finally we consider the zero limit of the noise, hoping that the relations survive, 
which is an approach similar to, e.g., \cite{fz, fl, k07}.
Although this limit can be difficult to study theoretically,
the derived relations can be studied numerically 
to check whether or not, with sufficiently small noise, they are good approximations 
for the noiseless original system.

More specifically, by adapting the method in \cite{hs05, hs06},
we here derive the FRR of a randomly perturbed GOY shell model
and the FRR of the velocity Fourier modes in the Eulerian frame 
of the randomly perturbed incompressible Navier-Stokes equations. 
We take the following steps:
(i) we add Gaussian white noise
to the equation of each shell variable and each velocity Fourier mode;
(ii) adapting the Harada-Sasa argument, we derive formally the FRR for these randomly perturbed
  shell variables and the velocity Fourier modes;
(iii) we consider numerically how small the noise should be
so that the randomly perturbed system recovers the noiseless system;
(iv) we numerically check whether or not the FRR derived in (ii) 
is consistent with the linear response function of the noiseless system
with the sufficiently small noise.
Concerning (iii) and (iv) above, the numerical analysis is carried out only for the shell model case
in this paper.

\textit{Derivation of the FRR}~We consider a version of the shell model \cite{sabra},
whose variables $u_j(t) ~(j = 0, \ldots, N)$ are complex numbers.
They are representatives of  the velocity Fourier modes of the incompressible Navier-Stokes equations
in the spherical shell $k_j \le |\vec{k}| < k_{j + 1}$ of the wavenumber space, where $k_j = k_0 2^{j}$.
The equation of $u_j(t)$ with the complex-number Gaussian white noise $\xi_j(t)$ is
\begin{eqnarray}
\mbox{$\frac{d}{d t}$} u_j(t)
 = \Lambda_j(t)	- \nu k_j^2 u_j(t) + \xi_j(t) + f^{(p)}_j(t),
  \label{shell}
\end{eqnarray}
where $\Lambda_j(t)$ includes the nonlinear term and
the deterministic
large-scale forcing $F_j(t)$ to keep the system statistically steady:
$\Lambda_j(t) = i (   k_j u_{j + 2}(t) u^*_{j + 1}(t)
       - \mbox{$\frac{1}{2}$} k_{j - 1} u_{j + 1}(t) u^*_{j - 1}(t)
       + \mbox{$\frac{1}{2}$} k_{j - 2} u_{j - 1}(t) u_{j - 2}(t)) + F_j(t)$.
Here $^*$ denotes the complex conjugation.
In Eq.~(\ref{shell}), $\nu$ models the kinematic viscosity and the noise $\xi_j(t)$ has
the mean and covariance
\begin{eqnarray}
\langle \xi_j(t) \rangle = 0, \quad 
\langle \xi_j(t) \xi^*_l(s) \rangle = 2 \sigma_j^2 T \delta_{jl} \delta(t - s),
\label{covari}
\end{eqnarray}
where $T$ is the strength of the noise which we here call ``temperature''.
We later compare numerically the FRR for small $T$ with that of the shell model without 
the noise.
The last term of Eq.~(\ref{shell}), $f^{(p)}_j(t)$,
is the infinitesimal probe force by which we define the linear response function.

To obtain an expression of the linear response function, 
we follow the Onsager-Machlup path-integral approach \cite{hs06}. 
The starting point is the probability functional of the
Brownian paths ${\xi_j(t)} ~(j = 0, \ldots, N)$ from time $t_0$ to $t$, 
$ P(\vec{\xi}, t| \vec{\xi}_0, t_0) 
= 
\int_{(\vec{\xi}_0, t_0)}^{(\vec{\xi}, t)} D[\vec{\xi}] 
\exp\left[
     -\mbox{$\frac{1}{2}$}
     \sum_{j = 0}^{N}
     \int_{t_0}^{t}
     \mbox{$\frac{|\xi_j(s)|^2}{\sigma_j^2 T}$}
     ds
    \right]$.
Change of variables from $\vec{\xi} = (\xi_0, \ldots, \xi_N)$ to $\vec{u}=(u_0, \ldots, u_N)$ 
yields the path-integral representation of the transition probability as
\begin{eqnarray}
&& P(\vec{u}, t|\vec{u}_0, t_0) =
 \int_{(\vec{u}_0, t_0)}^{(\vec{u}, t)} D[\vec{u}] 
    \exp\bigg\{
       -\mbox{$\frac{1}{2}$}
       \sum_{l = 0}^{N}
       \int_{t_0}^{t} ds \nonumber \\
&&       \bigg[
       \mbox{$\frac{1}{\sigma_l^2 T}$}
       |\dot{u}_l(s) - \Lambda_l(s) 
   + \nu k_l^2 u_l(s)  - f^{(p)}_l(s)|^2 \nonumber \\
&&       +
       \mbox{$\frac{\partial}{\partial u_l}$}
       (\Lambda_l(s) 
   - \nu k_l^2 u_l(s)  + f^{(p)}_l(s))       
       \bigg] \bigg\}.
\label{pu}       
\end{eqnarray}
The last divergence term can be interpreted as contribution from 
the Jacobian \cite{g77}. 
By linearizing Eq.~(\ref{pu}) in regard to $f^{(p)}_l$, we obtain
an expression of the ensemble average, $\langle u_j(t) \rangle_p$.
The mean linear response function $G^{(T)}_{jl}$ can be then
written as
\begin{eqnarray}
G^{(T)}_{jl}(t - s) &=&    \frac{\delta \langle u_j(t) \rangle_p}
        {\delta f_l^{(p)}(s)}
  = \mbox{$\frac{1}{2\sigma^2_l T}$}
  \bigg[
   \langle \dot{u}_l^*(s)  u_j(t) \rangle \nonumber \\
&&   - \langle \Lambda_l^*(s) u_j(t) \rangle
   + \nu k_l^2 \langle u_l^*(s) u_j(t) \rangle
  \bigg].
\label{Hjl}  
\end{eqnarray}
We denote the most right-hand side of Eq.~(\ref{Hjl}) by $H^{(T)}_{jl}(t - s)$.    
Here $\langle \cdot \rangle$ represents the ensemble average taken in the absence of
the probe force.
For the diagonal part $G^{(T)}_{jj}$, we can simplify the expression
by using the causality of the response function and the temporal symmetry 
of the autocorrelation function, as
\begin{eqnarray}
 G^{(T)}_{jj}(t - s) &=& 
   \mbox{$\frac{1}{\sigma_j^2 T}$}
   \bigg\{
   \nu k_j^2 C^{(T)}_{jj}(t - s) \nonumber \\
&&  - \mbox{$\frac{1}{2}$}
  \big[
  \langle \Lambda_j^*(t) u_j(s) \rangle
   +
  \langle \Lambda_j^*(s) u_j(t) \rangle   
  \big]\bigg\},
\label{diag}  
\end{eqnarray}
where
$C_{jj}^{(T)}(t - s) = \langle u_j(t) u^*_j(s) \rangle$ is 
the autocorrelation function. Equation (\ref{diag}) is the main FRR result
of this paper, which we study numerically below.

It is straightforward to extend the above argument to
the case of the three-dimensional incompressible Navier-Stokes equations
in a periodic cube, which are written in terms of the velocity Fourier
coefficients, 
$(\hat{u}_1(\vec{k}, t),\, \hat{u}_2(\vec{k}, t),\, \hat{u}_3(\vec{k}, t))$,
as
$ \mbox{$\frac{d}{d t}$} \hat{u}_a(\vec{k}, t)
 = - i \sum_{b, c = 1}^3 k_b( \delta_{ac} - \mbox{$\frac{k_a k_c}{k^2}$})
     \sum_{\stackrel{\vec{p}, \vec{q}}{\vec{p} + \vec{q} = \vec{k}}}
     \hat{u}_b(\vec{p}, t)
     \hat{u}_c(\vec{q}, t) 
 + \hat{F}_a(\vec{k}, t) 
   - \nu k^2 \hat{u}_a(\vec{k}, t).$
Here $k = |\vec{k}|$ and 
we assume that the deterministic large-scale forcing 
$\hat{F}_a$ is solenoidal,
and that the number of the Fourier coefficients is finite. 
Due to the incompressibility $\vec{k} \cdot \hat{\vec{u}}(\vec{k}, t) = 0$,
$\hat{\vec{u}}(\vec{k}, t)$ has only two independent components,
which we express as 
$\hat{\vec{u}}(\vec{k}, t) = \hat{u}_\varphi(\vec{k}, t) \vec{e}_\varphi + \hat{u}_\theta(\vec{k}, t) \vec{e}_\theta$ \cite{anglememo}.
To the equations of the two components,
we add the probe force $(f^{(p)}_\varphi, f^{(p)}_\theta)$ and
the Langevin noise $(\xi_\varphi, \xi_\theta)$  satisfying
$\langle \xi_\alpha(\vec{k}, t) \xi_\beta(\vec{q}, s) \rangle =
 2 \sigma^2(k) T \delta_{\alpha \beta} \delta_{\vec{k}, -\vec{q}}\delta(t - s)$
with the indices $\alpha, \beta = \varphi, \theta$.
The mean linear response function in the Navier-Stokes case is expressed
as
\begin{eqnarray}
 G^{(T)}_{\alpha \beta}(\vec{k}, t|\vec{q}, s)
  =
 \frac{\delta \langle \hat{u}_\alpha(\vec{k}, t) \rangle_p}
         {\delta f^{(p)}_\beta(\vec{q}, s)}
  = \mbox{$\frac{1}{2\sigma(k)^2 T}$}
    [  \langle \dot{\hat{u}}_\beta^*(\vec{q}, s) \hat{u}_\alpha(\vec{k}, t)\rangle \nonumber \\
 - \langle \Lambda^*_\beta(\vec{q}, s) \hat{u}_\alpha(\vec{k}, t)\rangle   
     + \nu k^2 \langle \hat{u}_\beta^*(\vec{q}, s) \hat{u}_\alpha(\vec{k}, t)\rangle
    ].
\end{eqnarray}
Here $\Lambda_\alpha(\vec{k}, t)$ is the $\alpha$ component of 
the sum of the first and second terms on the right hand side of the Navier-Stokes equations, 
defined similarly as in the shell model case.
Further simplification can be made for the sum of the diagonal parts as
\begin{eqnarray}
&& \mbox{$\frac{1}{2}$}
 \sum_{\alpha = \varphi, \theta}
 \int G^{(T)}_{\alpha \alpha}(\vec{k},t | \vec{k}, s) \mbox{$\frac{d\Omega_k}{4\pi k^2}$}
  =
\mbox{$\frac{1}{\sigma(k)^2 T}$}
[
 \nu k^2 E(k; t - s) \nonumber \\
&& \hspace*{2cm}
 - \mbox{$\frac{1}{2}$}( {\mathcal T}(k; t, s) + {\mathcal T}(k; s, t) ) 
] 
\mbox{$\frac{1}{4\pi k^2}$},
\label{nsdiag}    
\end{eqnarray}
where the integral is over the surface of the sphere of radius $k$.
Here we assume isotropy of the 2nd-order tensors $\langle \hat{u}^*_c(\vec{k}, s) \hat{u}_a(\vec{k}, t)\rangle$
and $\langle \Lambda^*_c(\vec{k}, s) \hat{u}_a(\vec{k}, t)\rangle$
with respect to $\vec{k}$ and use the two-time energy spectrum function
$E(k; t - s) = \frac{1}{2}\int \sum_{a = 1}^3 \langle \hat{u}^*_a(\vec{k}, s) \hat{u}_a(\vec{k}, t) \rangle ~d\Omega_k$
and the two-time energy transfer function
$\mathcal T(k; t, s) = \frac{1}{2}\int \sum_{a = 1}^3 \langle \Lambda^*_a(\vec{k}, s) \hat{u}_a(\vec{k}, t) \rangle ~d\Omega_k$.

The FRRs of the diagonal parts, Eqs.~(\ref{diag}) and (\ref{nsdiag}),
have an interesting structure: deviation from the proportional relation
between the linear response function and the auto-correlation function
is ascribed to
the correlations between the nonlinear term and the velocity,
which is \emph{the nonlinear energy transfer} for the equal-time case.
This suggests that 
the energy transfer between scales, or the energy cascade, causes
the deviation. 
We observe also that 
the condition $H^{(T)}_{jj}(0) = 1$ is satisfied if the squared modulus
of each mode is in a statistically steady state.

\textit{Numerical analysis of the shell model}~
The shell model Eq.~(\ref{shell}) with the total 19 shell variables ($N = 18$)
is numerically solved to 
check whether or not
the expression of the response function $H^{(T)}_{jj}$ with the noise
(the right hand side of Eq.~(\ref{diag})) as $T \to 0$ approaches the one without the noise.
We use a forth-order Runge-Kutta scheme with the time step $\Delta t = 10^{-4}$.
The parameter values are 
$k_0 = 6.25\times10^{-2},~ F_j = 5\times 10^{-2}(1 + i)\delta_{j0},~ \nu = 1.66\times10^{-5}$,
yielding the shell-model analogue of the Taylor-microscale Reynolds number ${\rm Re}_\lambda = 3.9\times10^6$
and of the large-scale turnover time $\tau_L = 0.60 = 6000 \Delta t$ \cite{mp04}.

We use a common numerical method to directly measure 
$G^{(T)}_{jj}(t - s)$, without using the probe force,
by following difference between a pair of orbits, $\Delta u_j(t)$ with
one orbit being slightly displaced from the other at time $s$ 
by $\Delta u_j(s)$.
This yields $G^{(T)}_{jj}(t - s) = \langle \Delta u_j(t) \rangle / \Delta u_j(s)$ \cite{bdlv, mprv08}.
The pair share the same realization of the noise.
The value of the past displacement $\Delta u_j(s)$, taken here to be real,
is set to five percents of the standard deviation of the real part of $u_j$.
We start with twenty different random initial conditions,
where all the real and imaginary parts of $u_j$ 
are set by uniformly distributed random variable between $-1$ and $1$. We first discard data upto $t_1 =3.3\times10^4\tau_L$ as
initial transients and measure the correlation functions and the response functions from $t_1$ to $t=8.3\times10^5\tau_L$.
The number of samples in calculation of $G^{(T)}_{jj}$ is $5 \times 10^5$.
For the variance of the noise, we here report the result with a simple choice $\sigma_j^2 = 1$. 
We test two other $k$-dependent settings, $\sigma_j^2 = \nu k_j^2,\, \nu k_j^{4/3}$,
and find that the case $\sigma_j^2 = 1$ yields the fastest approach to the noiseless
system at temperature $T=10^{-4}$ as shown below.
Here we do not intend to study the system by varying the power of the wavenumber in the
variance in the framework of the renormalization-group analysis of turbulence \cite{smp98, blt04, mmm07}.
The expression $H^{(T)}_{jj}$ is obtained by calculating separately 
the three correlation functions on the right hand side of Eq.~(\ref{diag}).
\begin{figure}
\centerline{\includegraphics[scale = 0.60]{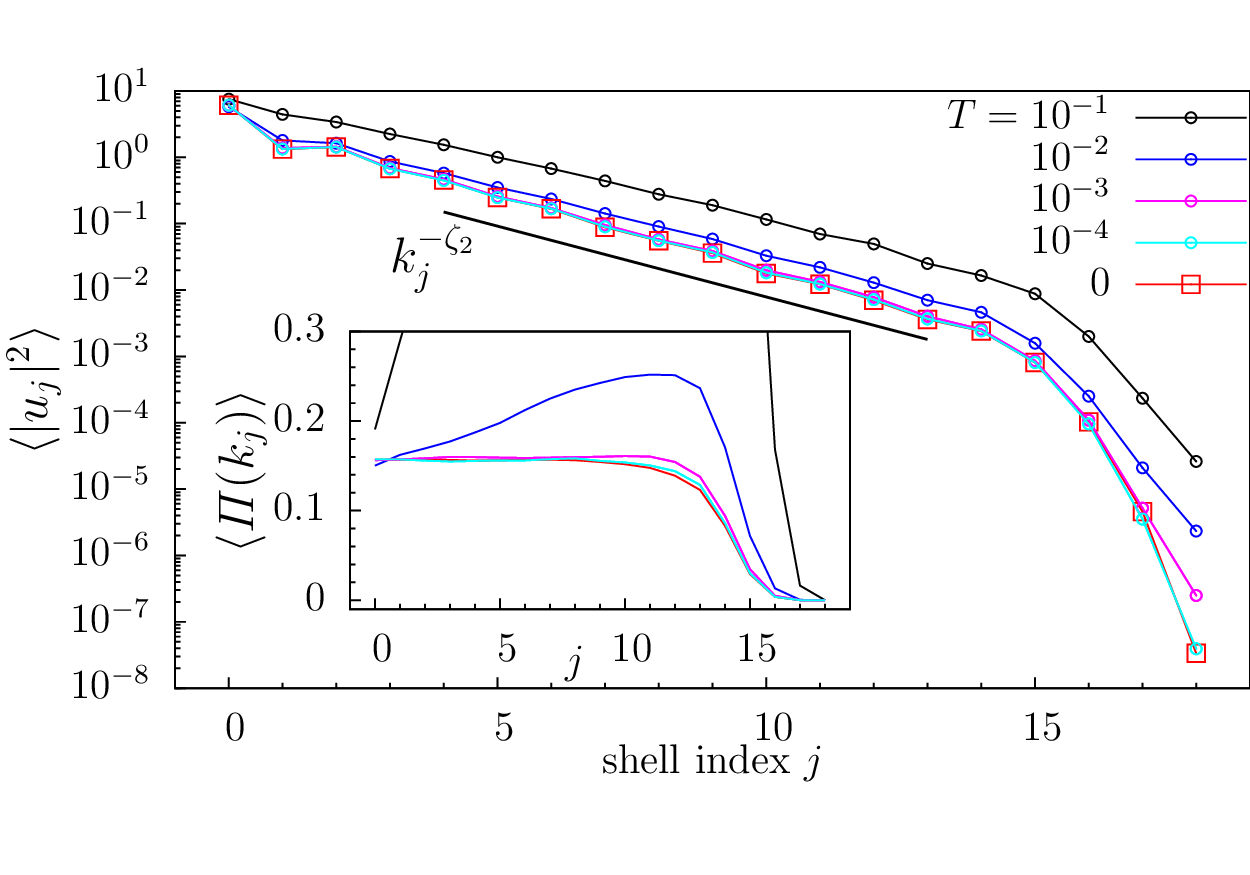}}
\caption{\label{spec} (Color online) 
Second-order moment of the absolute value of the shell-model variable $u_j(t)$ 
with and without the Gaussian white noise, as a function of the shell index.
Here $\sigma_j^2 = 1$ in Eq.~(\ref{covari}).
Inset: the energy flux function $\varPi(k_j)$ showing that
the constant-energy-flux structure is preserved for $T \le 10^{-3}$.}
\end{figure}

\begin{figure}
\centerline{\includegraphics[scale = 0.60]{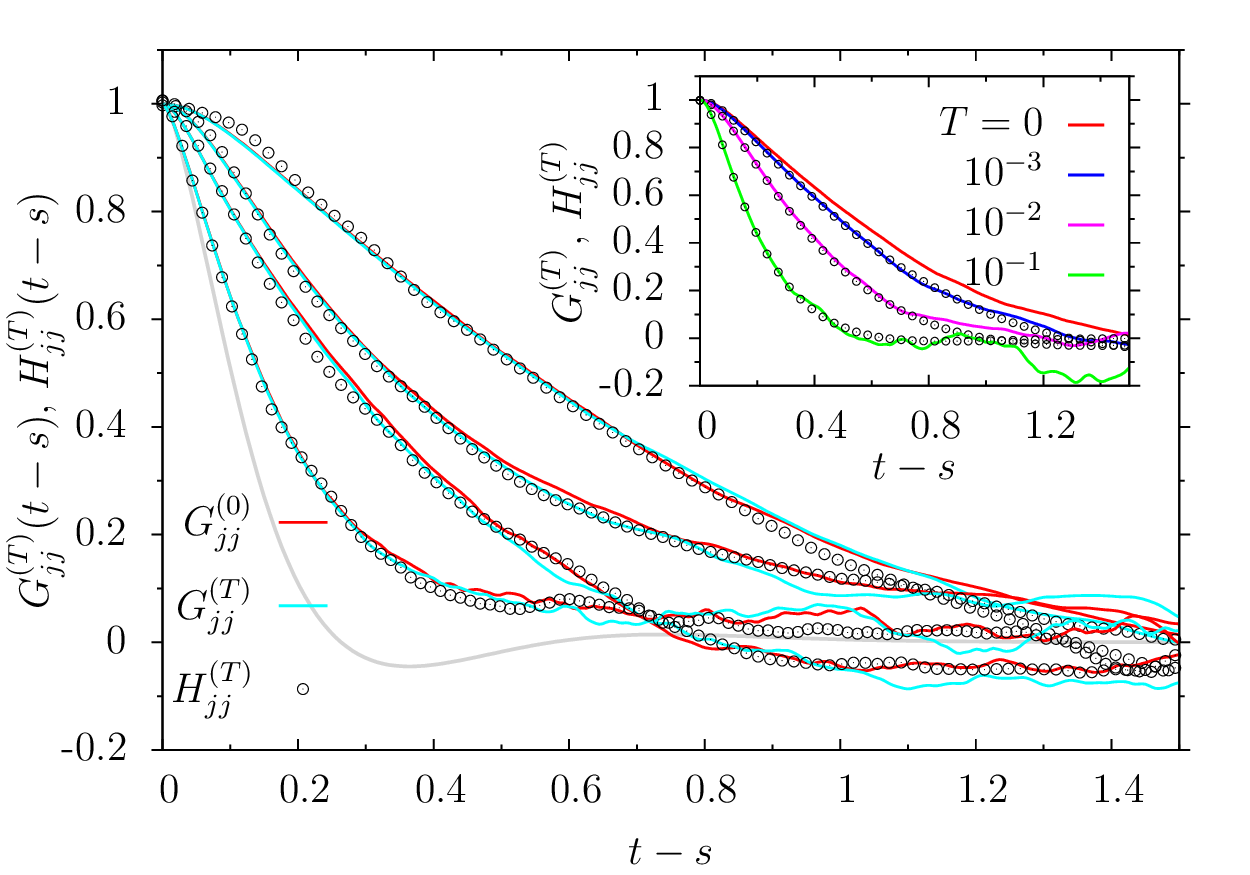}}    
\caption{\label{gh} (Color online)
Directly calculated response function of the shell model $G^{(0)}_{jj}$ (zero temperature),
$G^{(T)}_{jj}$ with $T = 10^{-4}$, 
and the FRR expression of the response function $H^{(T)}_{jj}$ with $T=10^{-4}$, the right hand side of Eq.~(\ref{diag}),
for the shell indices $j = 9, 10, 11$, and $12$ (from top to bottom). 
The gray curve is $C^{(0)}_{jj}(t - s)/ C^{(0)}_{jj}(0)$ for $j = 12$.
Inset: approach of $G^{(T)}_{jj}$ to $G^{(0)}_{jj}$ as $T \to 0$ for $j = 9$,
plotted with $H^{(T)}_{jj}$.}
\end{figure}

In Fig.~\ref{spec} we show 
the time-averaged second-order moment of $|u_j|$ 
exhibiting the inertial-range scaling $k_j^{-\zeta_2}~(\zeta_2 = 0.709)$ \cite{mp04}
and
the averaged energy flux function $\langle \varPi(k_j) \rangle = \langle \sum_{l=j + 1}^{N} {\rm Re}[(\Lambda_l - F_l) u^*_l]\rangle$ 
for various temperatures,
indicating that the basic statistics of the Langevin shell model, 
as $T \to 0$, become closer to those of the noiseless shell model ($T = 0$).
For the lowest temperature $T = 10^{-4}$, shown in Fig.~\ref{spec},
now let us demonstrate that the expression of the response function $H^{(T)}_{jj}$
agrees both with the directly measured response function $G^{(T)}_{jj}$ and
with that of the noiseless case $G^{(0)}_{jj}$ in Fig.~\ref{gh} \cite{ri-note}. 
Firstly, we observe that $G^{(T)}_{jj}$ approaches $G^{(0)}_{jj}$ as decreasing $T$,
which is displayed in the inset of Fig.~\ref{gh}. With $T = 10^{-4}$ the difference 
between $G^{(T)}_{jj}$ and $G^{(0)}_{jj}$ is less than a few percents 
for all the shell indices.
Secondly, the autocorrelation function $C^{(T)}_{jj}$ also approaches $C^{(0)}_{jj}$ 
for all the indices as well.
As observed in \cite{bdlv}, 
for any index $j$, $G^{(0)}_{jj}$ is not proportional to $C^{(0)}_{jj}$
(only the case for $j=12$ is presented in Fig.~\ref{gh}). 
Lastly, $H^{(T)}_{jj}$ agrees with $G^{(0)}_{jj}$ 
within a few percents for the shells $9 \le j \le 18$ covering
from the middle of the inertial range to the end of the dissipation range.
Four of these shells, $9 \le j \le 12$, are presented in Fig.~\ref{gh}.
Note also that $H^{(T)}_{jj}$ agrees with $G^{(T)}_{jj}$ for the higher temperature 
cases, even though $G^{(T)}_{jj}$ is distinctly different from $G^{(0)}_{jj}$ 
as shown in the inset of Fig.~\ref{gh}. 

\begin{figure}
\centerline{\includegraphics[scale = 0.50]{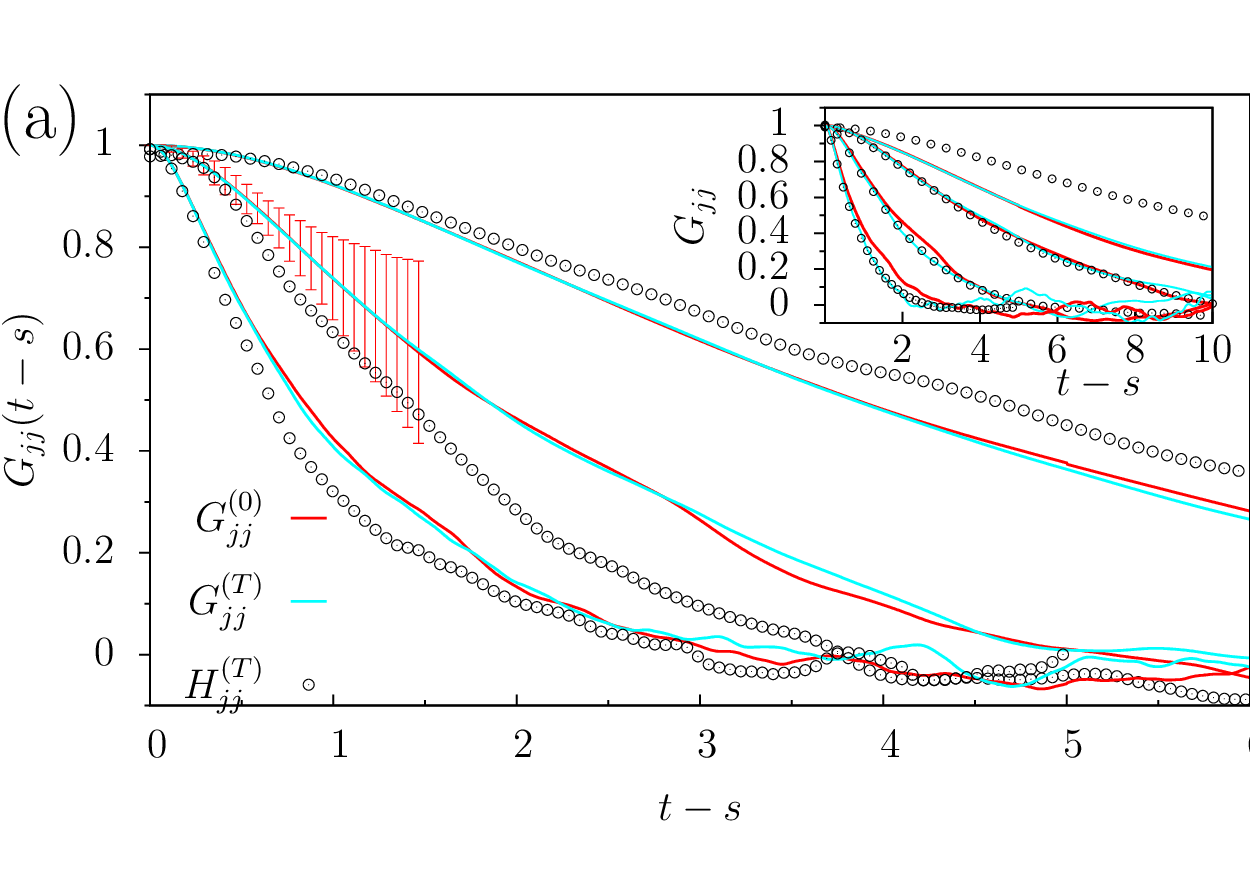}}
\centerline{%
\includegraphics[scale = 0.35]{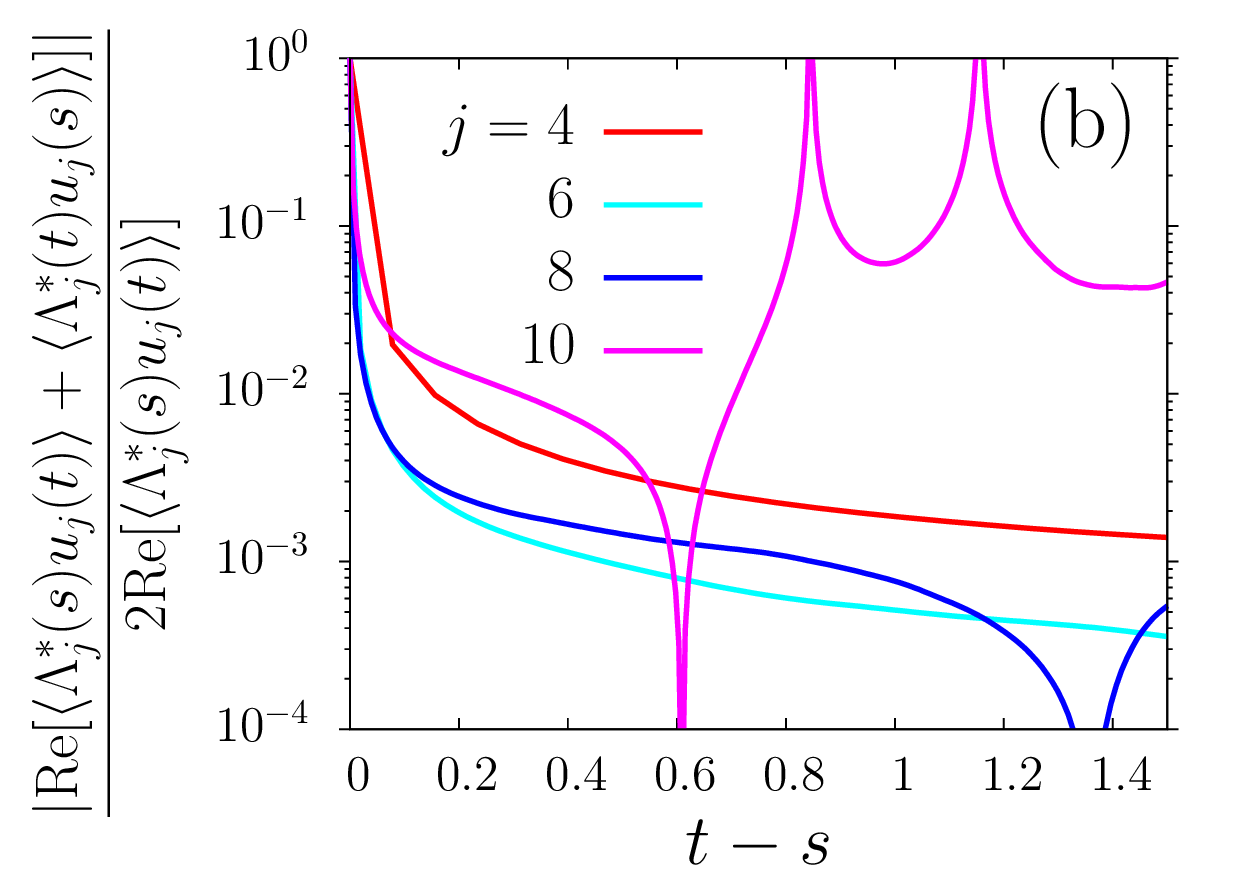}
\includegraphics[scale = 0.35]{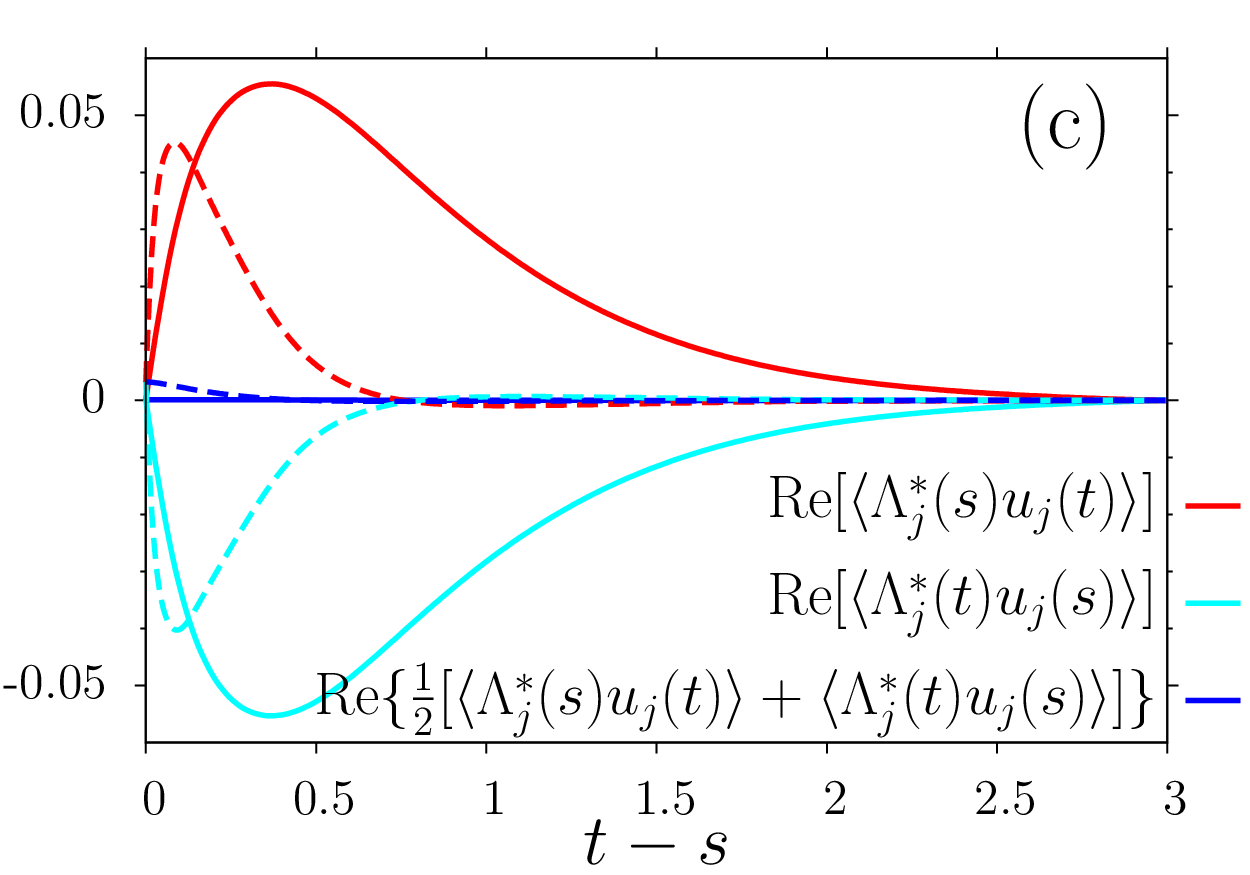}
 } 
\caption{\label{dis}(Color online) Numerical difficulty of the FRR. 
(a) Discrepancy between $H^{(T)}_{jj}$ and $G^{(0)}_{jj}$ 
for $j = 4, 6$, and $8$ (from top to bottom) with $T = 10^{-4}$;
The errorbars of $G^{(0)}_{jj}$ are shown for the index $j = 6$ up to $t - s= 1.5$.  
Inset: same as the outset but for $j = 2, 4, 6$, and $8$ with $T=10^{-3}$. 
 (b) Cancellation 
 between the real parts of $\langle \Lambda_j^*(s) u_j(t)\rangle$ and $\langle \Lambda_j^*(t) u_j(s)\rangle$ with $T = 10^{-4}$.
 (c) Symmetry of the real parts of $\langle \Lambda_j^*(s) u_j(t)\rangle$ and $\langle \Lambda_j^*(t) u_j(s)\rangle$ for $j = 8$ (solid) and $11$ (dashed)
     with $T = 10^{-4}$.
}
\end{figure}

For the shells  $0 \le j \le 8$, a discrepancy is observed for $T=10^{-4}$ 
as displayed in Fig.~\ref{dis}(a).
The half widths of the errorbars of $G^{(0)}_{jj}$ in Fig.~\ref{dis}(a)
correspond to the standard deviations among the $5 \times10^{5}$ samples of
$G^{(0)}_{jj}$.
Now we argue that this discrepancy between $H^{(T)}_{jj}$ and $G^{(0)}_{jj}$
observed for small shell indices is not physical but numerical. 
This is caused by cancellation of the significant digits in the sum of the last two 
terms in Eq.~(\ref{diag}). 
Empirically, if the sum of the two terms
${\rm Re}[\langle \Lambda_j^*(t) u_j(s)\rangle]$ and ${\rm Re}[\langle \Lambda_j^*(s) u_j(t)\rangle]$,
having opposite signs, loses more than two significant digits, agreement
between $H^{(T)}_{jj}$ and $G^{(0)}_{jj}$ is lost
as indicated in Fig.~\ref{dis}(b).
It is difficult to obtain third-order correlation functions of $u_j$ like
$\langle \Lambda_j^*(t) u_j(s)\rangle$ with three or more digit accuracy.
In fact, for small $j$'s, the energy transfer correlations $\langle \Lambda_j^*(t) u_j(s)\rangle$ and 
$\langle \Lambda_j^*(s) u_j(t)\rangle$ become increasingly symmetric with the horizontal axis
except around the origin $t=s$ as shown in Fig.~\ref{dis}(c), 
being a structure likely in common to the Navier-Stokes case.
Nevertheless, this symmetry is weakened with a larger 
temperature $T=10^{-3}$ (since the noise breaks the time-reflection symmetry) 
and a better agreement is obtained for $j = 4$ as in the inset of Fig.~\ref{dis}(a).

\textit{Concluding remarks}~We derived formally
the FRR of a statistically steady turbulent state of the 
shell model and the incompressible Navier-Stokes equations with the Langevin
noise by using the method of \cite{hs05, hs06}. 
For the shell model case, as decreasing the amplitude of the noise,
we demonstrated numerically that for the intermediate and small scales 
the derived FRR expression of the linear 
response function is indeed consistent with that of the noiseless
shell model. We consider the discrepancy observed in the large scales 
as caused by the limited accuracy of the statistical quantities. 
Our conclusion is that for all the shells the FRR, Eq.~(\ref{diag}), as $T \to 0$, 
converges to the response function of the shell model without the noise.
For the Navier-Stokes case, our preliminary numerical result on two-dimensional inverse-cascade
turbulence with a feasible averaging time indicates that Eq.~(\ref{nsdiag}) for small $T$ is a good approximation 
of the response function of the noiseless system. We encounter numerical difficulties similar to the
shell model case.
A numerical assessment of the FRR, Eq.~(\ref{nsdiag}), will be reported elsewhere.
Regarding the intermittency, it does affect each term in the FRR. However, the FRR on the whole
remains unaffected. This suggests that the FRR may be a bridge relation of the intermittency
or dynamic multiscaling (see, e.g., \cite{bct11, pmpp11}) among the 2nd-order, 3rd-order 
correlation functions and the response function if the intermittency is not susceptible to
the small Langevin noise. 
With this bridge relation, however, the wavenumber dependence of the integral time of 
the response function may not be described by the dominant multiscaling exponents of
the 3rd-order correlation functions, since the cancellation occurs as seen in Fig.\ref{dis}(c).
Future research directions to take further advantage of the vanishing
noise can be to develop a spectral closure approximation with the FRR response function 
obtained here
and to consider saddle-point solutions (instantons) in the integral 
Eq.~(\ref{pu}) as in \cite{k07, btc}, which may yield an interesting dynamical approach to turbulence.

\textit{Acknowledgments}~ We acknowledge delightful discussions
with So Kitsunezaki and Shin-ichi Sasa and the support by Grants-in-Aid for Scientific Research 
(C) No.~21540388 and (C) No.~24540404 from JSPS. 
A part of the numerical calculation was done on the HPC facilities at RIMS, Kyoto University.


\begin{thebibliography}{99}
\bibitem{anglememo}
In the decomposition
the orthonomal vectors $\vec{e}_\varphi$ and $\vec{e}_\theta$	are associated with
respectively the azimuthal angle ($0 \le \varphi \le 2\pi$) and the polar
angle ($0 \le \theta \le \pi$, defined as the angle from the positive part of the $k_z$ axis) 
for a given $\vec{k}$ in the spherical coordinates of the three dimensional wavenumber space.
\bibitem{ri-note}
Hereafter,
we consider only the real parts of $G^{(T)}_{jj}(t - s)$
and $H^{(T)}_{jj}(t - s)$ since
the absolute values of their imaginary parts
are smaller than  $0.12$ for $0 \le t - s \le 1.5$ for the shells shown in Fig.~\ref{gh}.
\end{thebibliography}
\end{document}